\documentclass[twocolumn,english,aps,prb,showpacs,superscriptaddress,amssymb,amsfonts]{revtex4}
\usepackage[T1]{fontenc}
\usepackage[latin9]{inputenc}
\usepackage{amsmath}
\usepackage{epstopdf}
\usepackage{bm}
\usepackage{graphicx}
\usepackage{amssymb}
\usepackage{color}
\usepackage{enumerate}

\makeatletter
\newcommand{\be}{\begin{equation}}
\newcommand{\ee}{\end{equation}}                  
\newcommand{\bea}{\begin{eqnarray}}
\newcommand{\eea}{\end{eqnarray}}
\newcommand{\beas}{\begin{eqnarray*}}
\newcommand{\eeas}{\end{eqnarray*}}

\newcommand{\tr}{\textrm{tr}}

\makeatother

\makeatother

\usepackage{babel}

\makeatother

\usepackage{babel}

\begin{document}
 
 \title{Entanglement and the Sign Structure of Quantum States}

\author{Tarun Grover}
\affiliation{Kavli Institute for Theoretical Physics, University of California, Santa Barbara, CA 93106, USA}

\author{Matthew P. A. Fisher}
\affiliation{Kavli Institute for Theoretical Physics, University of California, Santa Barbara, CA 93106, USA}
\affiliation{Department of Physics, University of California, Santa Barbara, CA 93106, USA}

%

\begin{abstract}
Many body quantum eigenstates of generic Hamiltonians at finite energy density  typically satisfy the ``volume law'' of entanglement entropy: the von Neumann entanglement entropy and the Renyi entropies for a subregion scale in proportion to its volume. Here we provide a connection between the volume law and the sign structure of eigenstates. In particular, we ask the question: can a  positive wavefunction support a volume law entanglement? Remarkably, we find that a typical random \textit{positive} wavefunction, exhibits a \textit{constant law} for Renyi entanglement entropies $S_n$ for $n>1$, despite arbitrary large amplitude fluctuations. We also provide evidence that the modulus of the finite energy density eigenstates of generic local Hamiltonians show similar behavior.
\end{abstract}

\maketitle

\section{Introduction} \label{sec:intro}
The intricate sign structure of quantum states plays an important role in fields as disparate as quantum chaos \cite{stratt1979, blum2002}, quantum Monte Carlo simulations \cite{marshall1955, perron}, as well as semiclassical quantum mechanics \cite{migdal}. In this paper, we point out that the sign structure is also important for  understanding the qualitative behavior of entanglement entropy in finite energy density eigenstates of a generic quantum many body system.

In sharp contrast to the celebrated `area law' scaling in the quantum ground states for the von Neumann entanglement entropy $S_{vN}$ \cite{srednicki1993, bombelli1986}, and the Renyi entanglement entropies $S_n$ with $S_{vN}, S_{n} \sim \ell_A^{d-1}$ in $d$ spatial dimensions, the finite energy density eigenstates instead typically satisfy a `volume law' scaling: $S_{vN}, S_n \sim \ell_A^{d}$, where $\ell_A$ is the linear size of the subregion for which entanglement is being calculated \cite{lubkin1978, lloyd1988, page1993}. For a generic, non-localized \cite{basko2006, huse2007, pal2011, bela, imbrie}, non-integrable systems, where the equivalence between microcanonical and canonical ensembles is expected to hold true vis-a-vis the `Eigenstate Thermalization Hypothesis' (ETH) \cite{deutsch1991, srednicki1994, rigol,deutsch2010}, the volume law scaling of entanglement is equivalent to the extensivity of thermal entropy. In particular, in these systems, the von Neumann entanglement entropy $S_{vN}$ of an eigenstate with energy density $e$ equals $s_{\text{thermal}}(e)V_A $, where $s_{\text{thermal}}(e)$ is the thermal entropy density and $V_A \sim \ell_A^d$ is the volume of region $A$. In this paper, we ask: what feature(s) in a quantum state are responsible for the volume law scaling? We provide evidence that the sign structure  of wavefunctions \cite{footnote:sign} is essential for obtaining volume law scaling for Renyi entanglement entropies $S_n$ for $n > 1$. In particular, we show that a generic \textit{positive} wavefunction in the Hilbert space, despite arbitrary large amplitude fluctuations, typically only shows a \textit{constant law}: $S_n \sim s_n$ for $n>1$, where $s_n$ are finite positive numbers. We also provide evidence that the same holds true for the modulus of finite energy density eigenstates of  local Hamiltonians.

At a heuristic level, the aforementioned volume law entanglement for excited states results from the random structure of eigenstates at finite energy density, which necessitates an $O(e^{s_{\textrm{thermal}} \ell_A^d})$ number of eigenvectors to faithfully represent the reduced density matrix for a region of size $\ell_A$, where $s_{\textrm{thermal}}$ is the corresponding entropy density. This is in contrast to the ground state wavefunctions, which typically have a much more `rigid' structure, thus typically requiring a much smaller number $\sim O(e^{\ell_A^{d-1}})$ of eigenvectors. This motivates us to explore the concept of ensemble of wavefunctions, which will be important for our discussion throughout. Specifically, consider the set of wavefunctions of the form

\be 
|\psi \rangle = \sum_C \psi(C) \,|C\rangle \label{eq:ensemble}  ,
\ee
where $|C\rangle$ is a basis vector in a certain local (i.e. real-space) basis and $\psi(C)$ are picked from a specific random distribution subject to the normalization condition $\sum_{C} |\psi(C)|^2 = 1$. Given such an ensemble, one may ask what is the average entanglement entropies $\langle S_n\rangle,\langle S_{vN}\rangle$? As shown several decades ago by Lubkin \cite{lubkin1978}, if $\psi(C)$ are random, real or complex numbers with respect to unitarily invariant Haar measure (i.e. the vector ${\vec\psi(C)}$ is distributed uniformly over a sphere of the size of the total Hilbert space), then $\langle S_{vN}\rangle, \langle S_n\rangle$ are maximal: $\langle S_{vN} \rangle = \langle S_n \rangle = \ln(|\mathcal{H}_A|) \sim V_A$, where $|\mathcal{H}_A|$ is the size of the Hilbert space for region $A$ while $\overline{A}$ denotes complement of subregion $A$, and we have assumed that the ratio $\frac{V_A}{ V_{\overline{A}}} < 1$ while both $V_A, V_{\overline{A}} \rightarrow \infty$. Due to ETH, an eigenstate of a lattice model at `infinite temperature' (i.e. at an energy density $e$ such that $\frac{\partial{s_{\text{thermal}}}}{\partial{e}} = 0$) also satisfies $S_{vN} = S_n = \ln(|\mathcal{H}_A|) \sim V_A$ and therefore, in this respect, resembles a typical member of the ensemble in Eqn.\ref{eq:ensemble}.

In this paper, we develop a relation between random ensembles and entanglement with an eye on the sign structure of many-body eigenstates. Does any arbitrary random ensemble yields a volume law entanglement, or does one require a more specific structure to the states comprising the ensemble? For example, as discussed in Ref.\cite{grover2014}, a `sign-random' wavefunction, where $\psi(C) = \pm 1$ with equal probability, recovers the full infinite-temperature entanglement entropy,  $\langle S_n \rangle = \ln(|\mathcal{H}_A|)$, despite no fluctuations in the amplitude $|\psi(C)|$. This motivates us to ask: would a random ensemble where the wavefunction is allowed to fluctuate in amplitude, but not in its sign, show a volume law entanglement? A naive guess is that this is indeed the case -- one can clearly construct wavefunctions which are positive in a local basis and have volume law $S_n$ for arbitrary $n$. For example, consider a `long-range triplet' state for a spin-1/2 system: $|\psi_{\textrm{LRT}}\rangle = \prod_{i} \left(| \uparrow\rangle_i |\downarrow\rangle_{j(i)} + |\downarrow\rangle_{j(i)} |\uparrow\rangle_i\right)$ where $j(i)$ denotes the triplet partner of the $i$'th spin and is chosen so that the distance $|i - j(i)|$ is of the order of the total system size for each $i$. Such a state can be easily demonstrated to have a volume law $S_n$ for all $n$. Surprisingly, our analysis of random positive ensembles shows that this naive caricature of volume law wavefunctions is  misleading: on average, positive states show a \textit{constant law} Renyi entanglement entropy for Renyi index $n>1$ whose magnitude does not depend on the size of the Hilbert space in region $A$. Therefore, states such as $|\psi_{\textrm{LRT}}\rangle$ which are positive and have a volume law entropy are extremely rare. We also study physical Hamiltonians, and find that they also agree with the aforementioned constant law when entanglement is computed for the modulus of finite energy density eigenstates.


\section{Average Entanglement Entropy for Random Positive Ensembles} \label{sec:averageS}

\begin{figure}
\begin{centering}
\includegraphics[scale=0.28]{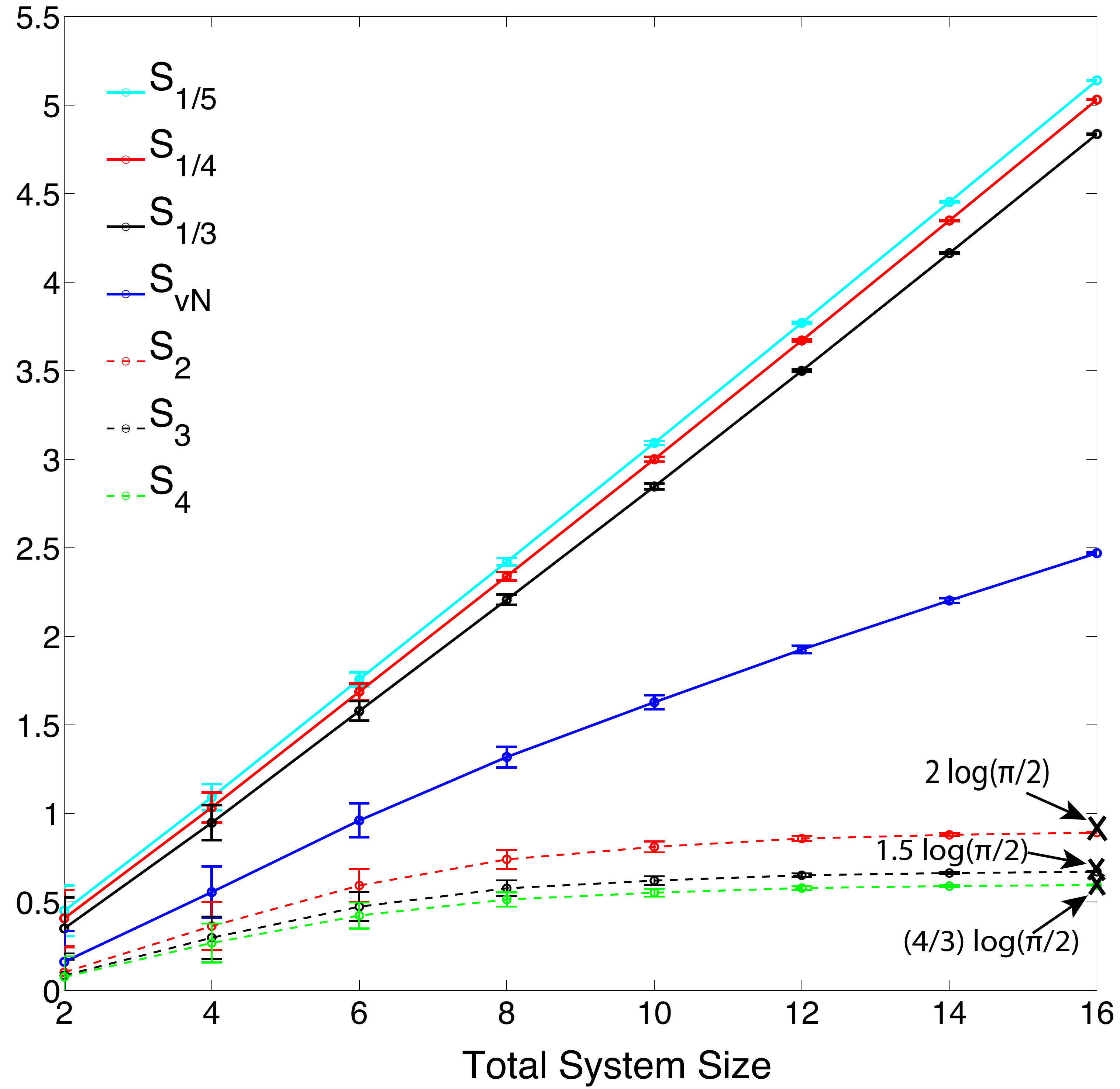}
\par\end{centering}
\caption{Average entanglement entropies $\langle S_n(\rho_A) \rangle$ corresponding to the Random Positive Ensemble. The region $A$ is half the total system size. $\langle S_n \rangle$ for $n>1$ ($n \leq 1$) shows a constant  law (volume law) whose values match with the analytical results for $ S_n(\langle\rho_A\rangle) $ for the same ensemble (see text for details). The `error bars' shown correspond to the variance of $S_n(\rho_A)$.} \label{fig:pos_rand}
\end{figure}

In order to understand the role of sign structure in generating entanglement, we ask: what is the average entanglement entropy of a  wavefunction which, in a given local basis, has only positive coefficients that are drawn from a specific random distribution? This may seem an ill-motivated question since a physical finite energy density state (e.g. a finite energy density eigenstate) will generically not be expandable with positive coefficients in a local basis. Furthermore, even if one restricts oneself to an ensemble of wavefunctions that have positive coefficients in a local basis, a change of basis will generically not maintain positiveness of the coefficients.  However, we find that the analysis of a random ensemble where the coefficients are positive in a fixed, chosen basis will lead to insights that are applicable more generally. We will examine the role of the choice of basis in Sec.\ref{sec:hamiltonian} where we study eigenstates of physical Hamiltonians.

Given an ensemble of wavefunctions, one may define at least three different measures of entanglement, depending on how one performs  the averaging: 
\bea 
S_n(\langle \rho_A \rangle) & = &  -\frac{1}{n-1} \ln(\,\tr\,  \langle \rho_A\rangle^n) , \\ 
S_n(\langle \tr\,\, \rho_A^n \rangle ) & = &   -\frac{1}{n-1} \ln(\, \langle \tr\, \rho_A^n \rangle), \\
\langle S_n( \rho_A ) \rangle & = &  -\frac{1}{n-1} \langle \ln \left( \tr \,\,\rho^n_A  \right) \rangle .\label{eq:defavg}
\eea
\noindent Here $n$ is any positive real number. Of these three, $\langle S( \rho_A ) \rangle$ is the easiest to interpret physically, and hardest to access analytically, while $S_n(\langle \rho_A \rangle) $ may seem the least physical, but is the one easiest to calculate. For brevity, we will sometimes denote $\langle S_n( \rho_A ) \rangle$ as $\langle S_n \rangle$.

Let us therefore consider a wavefunction,
\be 
\big|\,|\psi| \,\big\rangle = \sum_C |\psi(C)| \,|C\rangle \label{eq:rpe} ,
\ee
where $\{C\}$ spans the Hilbert space, and $\langle \psi|\psi\rangle = 1$. The random coefficients $\psi(C)$ are distributed uniformly on the sphere  $S^{|\mathcal{H}_A| |\mathcal{H}_{\overline{A}}|-1}$, or $S^{2|\mathcal{H}_A| |\mathcal{H}_{\overline{A}}|-1}$ depending on whether $\psi(C)$ are real or complex, the distinction between the two cases (i.e. real versus complex) being inconsequential for any of our results. Since the expansion coefficients $|\psi(C)|$ in the basis $|C\rangle$ are positive, we will refer to this ensemble as the ``Random Positive Ensemble'' (RPE). We will also study a less general case where the coefficients $\psi(C)$ in Eq.\ref{eq:rpe} are Slater determinants formed out of single-particle plane-wave states while the corresponding wavevectors are chosen at random from a uniform distribution over a Brillouin zone. 

A simple calculation (Appendix \ref{sec:rhoAspect}) shows that in general, $S_n(\langle \rho_A \rangle)$ for the wavefunction $\big|\,|\psi| \,\big\rangle $ in Eq.\ref{eq:rpe} is given by,

\be 
S_n(\langle \rho_A \rangle) = \frac{1}{1-n} \ln \left( (|\mathcal{H}_A|-1) \left( \frac{1-g}{|\mathcal{H}_A |}\right)^n +   \left(g+\frac{1-g}{|\mathcal{H}_A |}\right)^n\right) \label{eq:eeavgrho0}
\ee
where $g = \frac{\langle |\psi(C)|\rangle^2}{\langle |\psi(C)|^2 \rangle}$ and $|\mathcal{H}_A|$ denotes the size of the Hilbert space for subregion $A$. The expression for the von Neumann entropy is obtained by taking the limit $n \rightarrow 1$ in the above equation. For ease of presentation, below we will denote  $\ln(|\mathcal{H}_A|)$ by $\ell_A^d$, where $d$ is the spatial dimension, and $\ell_A$ is proportional to  the linear extend of the region $A$. The actual physical length differs from $\ell_A$ only by an $O(1)$ multiplicative factor that depends on the size of the local Hilbert space, which we ignore.



Let us consider the two aforementioned cases separately: 

(a) \underline{\textit{Random Positive Ensemble} (RPE)}: As mentioned above, this is the case for a state, $| \psi|$, where $\psi$ is an infinite temperature state which satisfies ETH. A simple calculation (Appendix \ref{sec:g}) shows that for this case, the parameter $g = 2/\pi$. This implies that in the asymptotic limit $V, V_A \rightarrow \infty$ with $V_A \leq V_{\overline{A}}$, one obtains,

\begin{equation}
S_n(\langle \rho_A \rangle) =  
\begin{cases}
\frac{n}{n-1} \ln \left(\frac{\pi}{2}\right) \,\, \text{if  } n > 1,\\
(1-\frac{2}{\pi}) \,\ell_A^d\,\, \text{if  } n = 1 ,\\
\ell_A^d\,\, \text{if  } n \leq 1 ,
\end{cases} \label{eq:eeavgrho1}
\end{equation}
with $S_1 \equiv S_{vN}$, the von Neumann entanglement entropy. Thus, all the Renyi entropies for $n > 1$ satisfy a $\textit{constant law}$, in sharp contrast to the ensemble of complex or real wavefunctions which satisfy a volume law with maximal coefficient: $S_n =  \ell_A^d$. At the same time, the asymptotic scaling of $S_n$ for $n < 1$ remains exactly the same as the one for the random complex/real ensemble while $S_{vN}$ displays a reduced volume law prefactor. 

The contrasting behavior for $n > 1$ and $n \leq 1$ signals a finite temperature phase transition in the (averaged) entanglement Hamiltonian, $H_\rho = -\ln(\langle \rho_A\rangle)$, for the positive ensemble. Indeed, $\langle \rho_A \rangle$ has $|\mathcal{H}_A|-1$ number of degenerate eigenvectors with a rather small eigenvalue of magnitude $\frac{(1-\frac{2}{\pi})}{|\mathcal{H}_A|}$, and a single eigenvector with a large eigenvalue $\frac{2}{\pi} + \frac{(1-\frac{2}{\pi})}{|\mathcal{H}_A|}$ which results in the phase transition at a temperature $1/n = 1$ for the entanglement Hamiltonian (see Appendix \ref{sec:rhoAspect} for details). 

We also calculated the two other measures of average entanglement for the RPE: $S_n(\langle \tr\,\, \rho_A^n \rangle )$ and the most physically relevant $\langle S_n( \rho_A ) \rangle$. As shown analytically in the Appendix \ref{sec:trrhoAn}, the result for $S(\langle \tr\,\, \rho_A^n \rangle )$ matches exactly with those for $S_n(\langle \rho_A \rangle)$ in Eqn.\ref{eq:eeavgrho1} . Finally, we numerically calculated $\langle S_n( \rho_A ) \rangle $ for the RPE for a total Hilbert space size $|\mathcal{H}_A| |\mathcal{H}_{\overline{A}}|$ up to $2^{16}$ (see Fig. 1) and find nearly perfect agreement with Eqn.\ref{eq:eeavgrho1}. Therefore, for the RPE, all three measures of entanglement $S_n(\langle \rho_A \rangle), S_n(\langle \tr\,\, \rho_A^n \rangle )$ and $\langle S_n( \rho_A ) \rangle$ agree with one  another.

\begin{figure}
\begin{centering}
\includegraphics[scale=0.36]{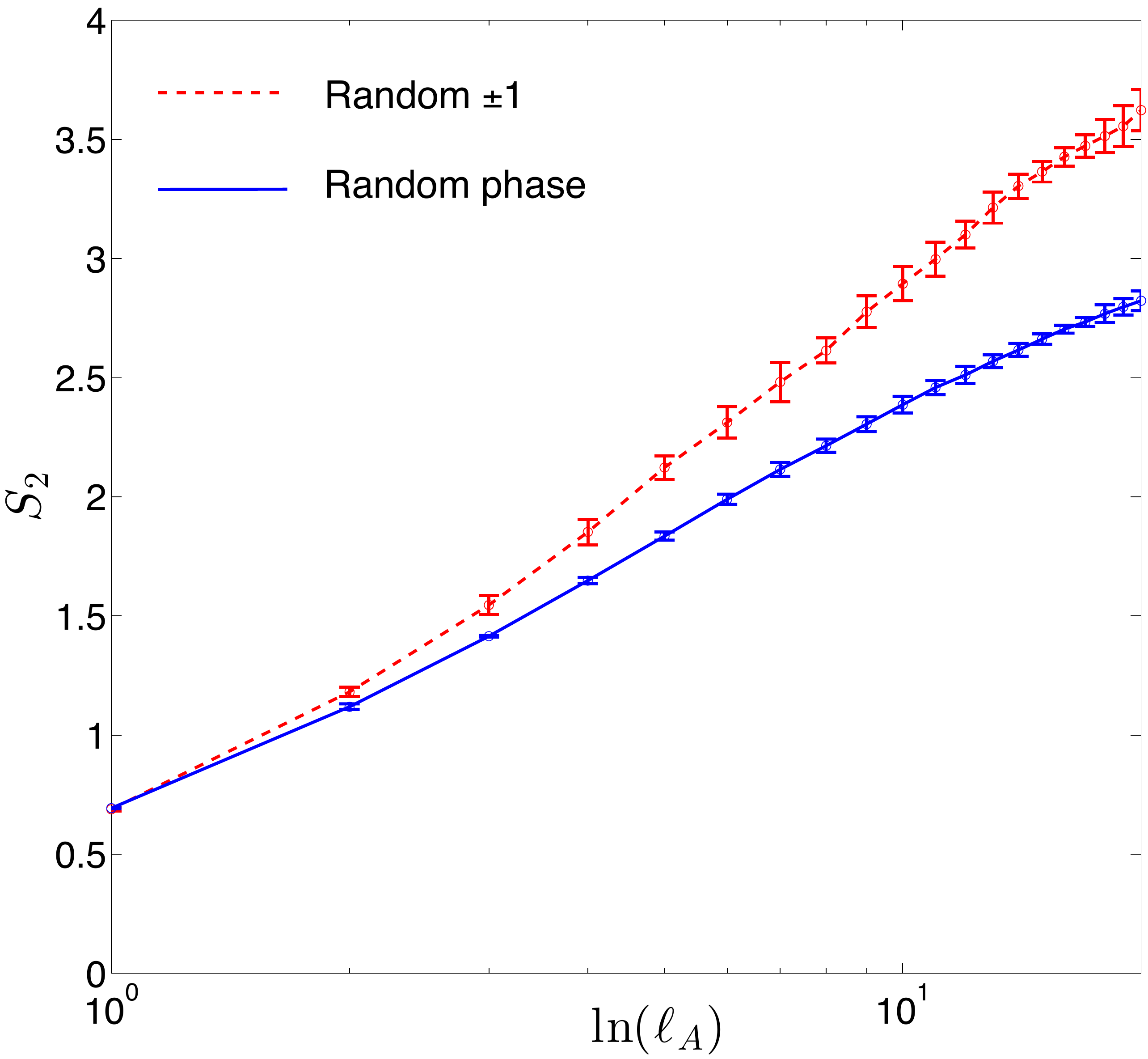}
\par\end{centering}
\caption{Renyi entanglement entropy $\langle S_2(\rho_A) \rangle$ corresponding to a wavefunction $\psi = |\det(M)|$ where $M$ is a matrix whose elements are plane wave states with random wavevectors (blue solid curve) or a matrix whose elements are $\pm 1$ with equal probability (red dashed curve). In the former case, we find  $S_2 \approx \frac{3}{4}\ln(\ell_A)$ while in the latter case,  $S_2 \approx \ln(\ell_A)$. The total system size is fixed at 60 sites, while $\ell_A$ varies from 1 to 20 sites. The error bars reflect the Monte Carlo sampling standard devitation, and not the actual variance of $S_2$ over the random ensemble.} \label{fig:freefermi}
\end{figure}

(b) \underline{\textit{$\psi(C) =$ random Slater determinant}}:
In this case, the original wavefunction is $\psi(C) = \det(e^{i \vec{k}_i.\vec{r}_j(C)})$, up to normalization, where the vector $\vec{k}$ is chosen from a uniform distribution over the 1D Brillouin zone, and the vector $\vec{r}(C)$ denotes the configuration $C$ in  real space (this is equivalent to choosing complex numbers of unit modulus with a uniformly distributed argument). Since $\psi$ corresponds to an integrable system, namely free fermions on 1D lattice, this case is non-generic, although still instructive. The results of Refs.\cite{storms2014, lai2014} imply that for a total system size of length $L$, the equality between the von Neumann entropy and thermal entropy for such a wavefunction holds only in the limit $\ell_A/L \rightarrow 0$, while $L \rightarrow \infty$, unlike the case of RPE where it holds as long as $\ell_A \leq L/2$, while $L \rightarrow \infty$).  

The numerical results for $\langle S_2( \rho_A ) \rangle$ corresponding to the wavefunction $\big|\,|\psi| \,\big\rangle$ in Eq.\ref{eq:rpe} are shown in Fig.2. These results are calculated using the quantum Monte Carlo sampling discussed in Ref.\cite{zhang2010}.  We find that $\langle S_2 \rangle \sim \alpha \ln(\ell_A)$, with $\alpha \sim \frac{3}{4}$, which is reminiscent of the $S_n \sim c \ln(\ell_A)$ for 1+1-d conformal field theories at zero temperature ($c$ is the central charge), although in contrast, the logarithmic scaling should hold in all dimensions since there is no notion of locality. Therefore, the positive random Slater determinant does not support a volume law entanglement either, although the entanglement is larger compared to the RPE discussed above. 

A partial understanding of the logarithmic scaling of $\langle S_2 \rangle$ is obtained by analytically calculating  $S_n(\langle \rho_A \rangle)$.  As is obvious from Eq.\ref{eq:eeavgrho0}, $S_n(\langle \rho_A \rangle)$ is independent of $\ell_A$, and depends only on $L$, the total system size. Therefore, unlike the case of RPE above where all three measures of entanglement (Eq.\ref{eq:defavg}) were asymptotically independent of the ratio $\ell_A/L$, here we don't expect $S_n(\langle \rho_A \rangle)$ to capture the full $\ell_A/L$ dependence of $\langle S_n( \rho_A ) \rangle$. Nonetheless, it may still capture the correct scaling behavior of  $\langle S_n( \rho_A ) \rangle$ when $\ell_A = r L$, with $r$ non-zero and fixed, so that there is only one scale in the problem. We have checked that this is indeed the case for a specific ensemble where $\psi(C)$ is a determinant of a matrix with random $\pm 1$ entries. We chose this particular ensemble, because the probability distribution function for the modulus of determinant for this ensemble was recently calculated in Ref.\cite{nguyen2014} (see also Ref.\cite{tao2006}) which allows us to calculate $S_n(\langle \rho_A \rangle)$ analytically via  Eq.\ref{eq:eeavgrho0}. We find  that the parameter $g$ in Eq.\ref{eq:eeavgrho0} scales as $g \sim 1/\sqrt{L}$ in contrast to the RPE, where it was a constant (see Appendix \ref{sec:g} for details).  Therefore, for this particular ensemble we find,


\begin{equation}
S_n(\langle \rho_A \rangle) =  
\begin{cases}
\ln \left(\ell_A\right) \,\, \text{if  } n > 1,\\
\\
\ell_A^d\,\, \text{if  } n \leq 1 ,
\end{cases} \label{eq:eeavgrho2}
\end{equation}

\noindent when $\ell_A = r L$ with $r$ non-zero and fixed.  As already hinted above, we find that the physically more  relevant $\langle S_2( \rho_A ) \rangle $ shows exactly the same scaling behavior, including the prefactor of unity for the logarithm: $\langle S_2( \rho_A ) \rangle
 = \ln(\ell_A)$, see Fig.\ref{fig:freefermi}. Even though the prefactor of the logarithm is slightly different than the case when $\psi(C) = \det(e^{i\vec{k}_i.\vec{r}_j(C)})$ with $\vec{k}$ random (1 instead of $\frac{3}{4}$), the qualitative behavior evidently remains unchanged. We expect the scaling $\langle S_n \rangle \propto \ln(\ell_A)$ when $n > 1$, and $\langle S_n \rangle \propto \ell^d_A$ when $n \leq 1$, to hold in general dimensions, akin to Eq.\ref{eq:eeavgrho2}. 
 
In passing, we mention that we also studied a case where only a fraction $f$ of $\vec{k}$ points are chosen randomly while the rest are contiguous. Not surprisingly, as $f \rightarrow 0$, the coefficient of the logarithm in the equation $\langle S_2 \rangle \propto \ln(\ell_A)$ for the wavefunction $|\psi|$ approaches $\frac{1}{4}$, since when all $k$ points are contiguous, the $|\det(e^{i \vec{k}_i.\vec{r}_j})|$ corresponds to the conformally invariant ground state of a 1D hard-core bosonic system \cite{lieb1961} whose entanglement entropy is $S_n = \frac{c}{6}\left( 1+ \frac{1}{n}\right)\ln(\ell_A)$ \cite{holzhey1994, cardy2004}.


\section{Relation to Physical Hamiltonians} \label{sec:hamiltonian}

\begin{figure}
\begin{centering}
\includegraphics[scale=0.4]{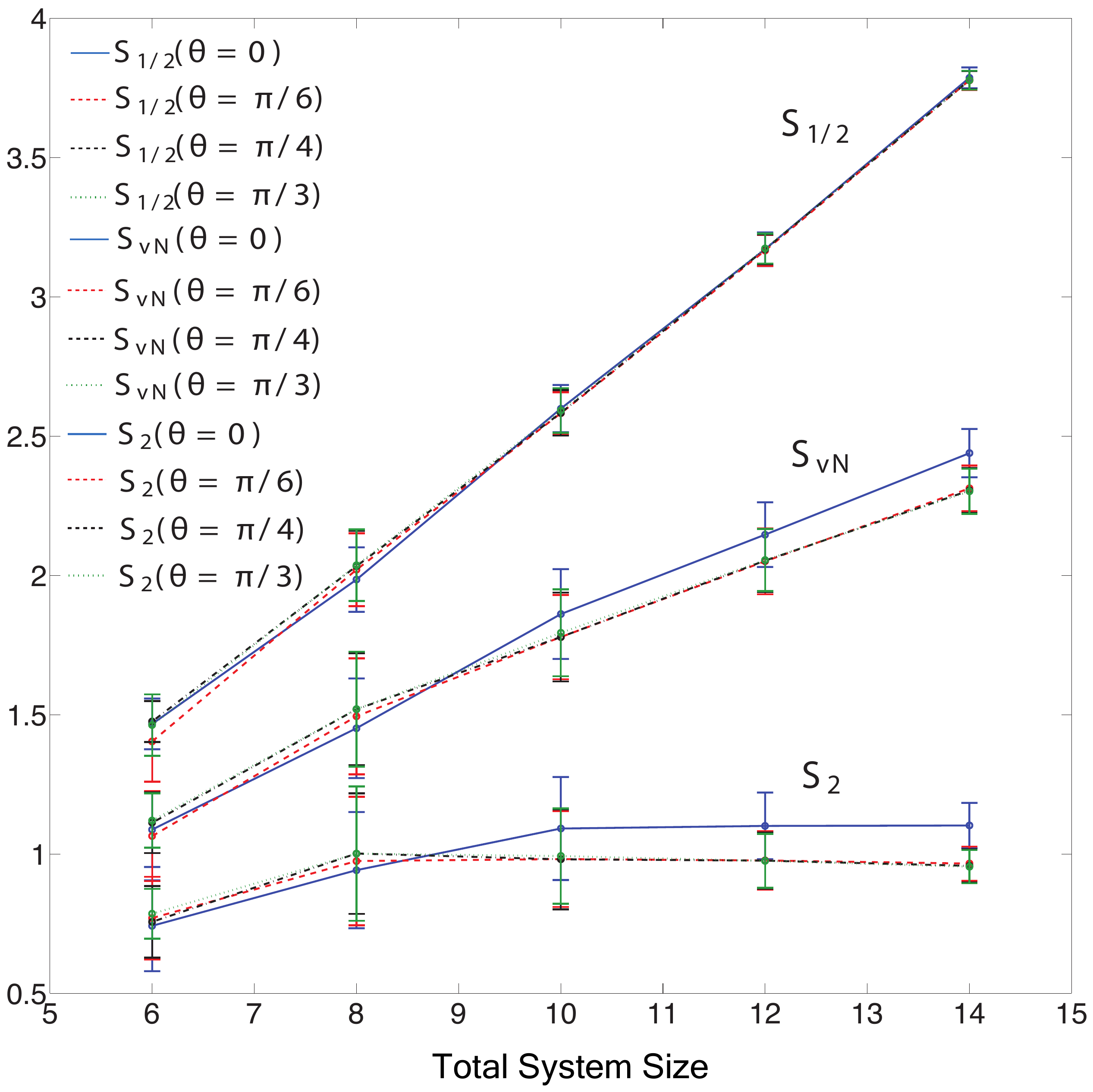}
\par\end{centering}
\caption{$\langle S_n \rangle$ for several $n$ corresponding to  the modulus of the eigenstates of the Hamiltonian in Eqn.\ref{eq:tfi}  close to the infinite temperature. The entanglement bipartition divides the total system into two equal halves. The plots for different $\theta$ correspond to the global rotation of the local basis by an angle $\theta$ along the $\hat{y}$ axis (recall that the entanglement for $|\psi|$ is basis-dependent). The qualitative behavior is found to be exactly same as that for the Random Positive Ensemble (RPE), as in Fig.\ref{fig:pos_rand}.} \label{fig:basis}
\end{figure}

The results of the previous section indicate that the sign structure is essential to obtain volume law entanglement for the Renyi entropies corresponding to a generic state in the Hilbert space. Here we provide further evidence for this statement by studying eigenstates of a non-integrable Hamiltonian. Specifically, consider the following Hamiltonian for a spin-1/2 chain:

\be 
H =  \Sigma_{i} \left(-\sigma^z_i \sigma^z_{i+1} + h_x \sigma^x_i + h_z  \sigma^z_i \right) \label{eq:tfi} ,
\ee
where the $\sigma$'s are spin-1/2 Pauli matrices and we impose periodic boundary conditions. We set $h_x = h_z = 1$; we verified that the qualitative features of our results remain true for other parameters as well as long as the Hamiltonian stays non-integrable. 

We diagonalized the above Hamiltonian for system sizes up to 12 sites and investigated eigenstates close to  infinite temperature by averaging over 1/8th of the total number of eigenstates around the part of the $S_{vN}(E)$ curve where $\frac{\partial S_{vN}}{\partial E}=0$, $E$ being the energy eigenvalue. Akin to the random ensembles studied in the previous section, we focus on  the entanglement structure of the modulus of these infinite temperature eigenstates. We first calculate entanglement in the $\sigma^z$ basis, and discuss the basis dependence in detail below. As shown in Fig.\ref{fig:basis}, we find clear evidence for a constant law for $\langle S_{n}\rangle$  when $n > 1$ and volume law for $n < 1$ and for $\langle S_{vN} \rangle$, akin to the RPE. This is in accordance with the fact that the Hamiltonian $H$ in Eq.\ref{eq:tfi} is non-integrable and is expected to satisfy ETH. The numerical values of the constant law are also very close to those found for the RPE.

As cautioned earlier, unlike the entanglement  for an actual eigenstate $\psi$, the entanglement corresponding its modulus $|\psi|$ in general depends on the choice of the local basis. Therefore, we next study the basis dependence of our results, again close to the infinite temperature. In particular, we consider the Hamiltonian 

\be
H '(\theta) = U^{\dagger}(\theta) H U(\theta) ,
\ee
where $U(\theta)$ denotes a global rotation of spins around the $\hat{y}$ axis by an angle $\theta$ with $H'(\theta = 0) \equiv H$. We obtain the eigenstates $\psi_\alpha$ of $H'(\theta)$ in the $\sigma^z$ basis, and study the entanglement entropies corresponding to $|\psi_\alpha|$. We find that both qualitatively and quantitatively, the results are rather insensitive to the choice of basis. This is not completely surprising -- assuming ETH holds, the coefficients $\psi(C)$ corresponding to the wavefunctions of $H$ are distributed uniformly over the sphere the size of Hilbert space, and a global rotation does not alter the random distribution.

We also studied the Renyi entropies of wavefunctions away from the infinite temperature. Our numerics indicate that the constant law for the Renyi entropies continues to hold {\it away from}  infinite temperature. This is also consistent with our quantum Monte Carlo results (not shown) for $\langle S_2(|\psi|) \rangle$, where we chose a variational wavefunction $\psi$ such that, when expanded in a local basis, it's sign structure is random while the amplitudes are not distributed uniformly on the sphere $S^{N_A N_B-1}$, thus mimicking a finite temperature state. As $T \rightarrow 0$, we expect that one recovers the area law entanglement for $|\psi|$ generically-- this is obvious for bosonic Hamiltonians whose ground state is positive in a local basis though we expect it to be true more generally (see also Ref.\cite{zhang2010}).

%


\section{Concluding Remarks}
Our main result is that the Renyi entropies $\langle S_n(|\psi|) \rangle$, $n > 1$, do not scale with volume, and instead show a constant law when $\psi$ is either a random wavefunction, or an eigenstate of a physical Hamiltonian close to infinite temperature. This is related to the fact that the off-diagonal elements of $\langle \rho_A(|\psi|) \rangle$ in any local basis are of the same magnitude as the diagonal elements. In contrast, the off-diagonal elements of $\langle \rho(\psi) \rangle$ are exponentially smaller in subsystem size compared to the diagonal elements leading to volume law Renyi entropies.

A slightly different perspective is obtained by noting that the Renyi entropies $S_n$ for integer $n>1$ may be decomposed into a `sign' and a `modulus' contribution \cite{zhang2010}. For example,

\be
S_2(\psi) = S_2(|\psi|)  + S^{\textrm{sign}}_{2}(\psi) ,
\ee

where $S_2(|\psi|)$ is the Renyi entropy corresponding to $|\psi|$ while $ S^{\textrm{sign}}_{2}(\psi)$ is defined via:

\be 
 e^{-S^{\textrm{sign}}_{2}(\psi)} = \sum_{C_1 C_2} \rho_{C_1,C_2} \textrm{sign}\left[ \psi(C_1) \psi(C_2) \psi(Sw_A C_1) \psi(Sw_A C_2)\right]
\ee

\noindent where $\sum_{C_1 C_2}$ denotes sum over configurations in two copies of the system, $\psi(Sw_A C_1)$ and $\psi(Sw_A C_2)$ are wavefunctions that are obtained by swapping the field configuration in subregion $A$ for $\psi(C_1)$ with those for $\psi(C_2)$ and $\rho_{C_1,C_2}$ is the probability density defined by  $\rho_{C_1,C_2}= |\psi(C_1)| |\psi(C_2)| |\psi(Sw_A C_1)| |\psi(Sw_A C_2)|$. Above, we have assumed that the wavefunction is real; the corresponding expression for complex wavefunctions  is very similar (see Ref. \cite{zhang2010}). At infinite temperature, all configurations are equally likely, and therefore one may approximate $S^{\textrm{sign}}_{2}(\psi)$ as $S_{2}(\textrm{sign}(\psi))$. Furthermore, assuming ETH holds,  $\textrm{sign}(\psi)$ will be completely random at  infinite temperature, and therefore $S^{\textrm{sign}}_{2}(\psi)$ equals the second Renyi entropy for the `sign-random' wavefunction discussed in Ref.\cite{grover2014}. For a given real-space basis vector in the Hilbert space, such a wavefunction takes values either +1 or -1 with equal probability. As shown in Ref.\cite{grover2014}, the Renyi entropy for a sign-random wavefunction is maximal i.e. when $V_A \leq V_{\overline{A}}$, $S_{2}(\textrm{sign}(\psi)) = \ln(\mathcal{H}_A) = \ell^d_A$, consistent with our detailed calculations which show that $S_2(|\psi|)$ doesn't contribute to the volume law entanglement \textit{at all} and as far as the contribution to the volume law entanglement is concerned, one may equate $S_2(\psi)$ with $S^{\textrm{sign}}_{2}(\psi)$. This discussion applies to $S_n$ for any integer $ n \geq 2$.

As a potential application of our results, one may consider writing down variational wavefunctions $\psi$ for highly excited states (e.g. the ground state of $(H-E)^2$ would be an excited state of $H$ with energy $E$). How might one verify that such wavefunctions have the correct entanglement structure? For a finite energy density eigenstate, calculating Renyi entropies $S_n(\psi)$ ($n \geq 2$) using Monte Carlo \cite{zhang2010} is extremely expensive from a computational standpoint, since one needs to calculate $\tr\,\left( \rho^n_A(\psi)\right)$ which scales as $e^{-V_A}$ where $V_A$ is the volume of region $A$. On the other hand, our results imply that $S_n(|\psi|)$ is straightforward to calculate since its computational complexity doesn't scale with the system size at all due to the constant law. Indeed, this is the reason that we were able to perform Monte Carlo calculations for some of the results presented in this paper. Therefore, $S_n(|\psi|)$ can provide an insight into the entanglement structure of such highly excited states while still being computationally accessible.

Our result is reminiscent of the relation between the number of nodes and the kinetic energy in elementary quantum mechanics -- typically, more nodes result in higher kinetic energy, and as we argued in the many-body context, higher entanglement entropy as well. This is not too surprising given that entropy and energy are directly related via $dE = TdS$ (recall that ETH implies that entanglement entropy equals the thermal entropy). We also note that ground states of bosonic systems are often nodeless in an appropriate local basis, which correlate with the fact that ground states typically do not exhibit volume law scaling of entanglement. Furthermore, as corroborated numerically in Ref.\cite{zhang2010}, even for systems with a Fermi surface which  show a multiplicative logarithm violation of area law, the modulus of the wavefunction only shows an area law entanglement. On that note, it will be interesting to explore the differences in the nodal structure for bosons and fermions in the excited states, and their manifestations in the corresponding entanglement structure.

We note that the essential role played by the random sign structure in obtaining the volume law also manifests itself in states that do not satisfy ETH. Consider a many-body localized phase where eigenstates obey an area law for the von Neumann entanglement entropy (and consequently area law Renyi entropies as well). As recently argued, there exist quasi-local unitary bases in which eigenstates can be expanded with positive coefficients \cite{huse2013, altman2012, abanin2013, swingle2013, ros2014}. This is consistent with the intuition developed in this paper that a volume law Renyi entropy indicates  that generically, there exist no local basis in which the wavefunction can be expanded with positive coefficients. Similar reasoning applies to the area law obeying ground states of systems that satisfy Marshall sign rule such as the ground states of the Heisenberg antiferromagnet on bipartite lattices.

In this paper we focused on wavefunction ensembles and eigenstates of local Hamiltonians to understand connection between quantum entanglement and the sign structure of quantum states. It might also be interesting to understand the role of sign structure in  quantum dynamics. An elementary insight along this direction follows from comparing the real time versus the imaginary time evolution of a quantum state. For a system that satisfies ETH, the real time evolution of a direct product state will eventually lead to a state whose entanglement entropy equals the thermal entropy \cite{srednicki1994}. In contrast, the \textit{imaginary} time evolution corresponds to projection onto the ground state wavefunction which would typically satisfy area-law entanglement (up to logarithmic corrections). This is reminiscent of the contrast between random complex ensemble and the positive random ensemble considered in this paper. We leave further exploration along this direction to the future.

{\bf Acknowledgements:}
We thank Leon Balents, Matthew Hastings and Patrick Hayden for stimulating conversations and Van Vu for a correspondence. This research was supported in part by the National Science Foundation, under Grants No. NSF PHY11-25915 and DMR-14-04230 (M.P.A. Fisher), and by the Caltech Institute of Quantum Information
and Matter, an NSF Physics Frontiers Center with support of
the Gordon and Betty Moore Foundation (M.P.A.F.).
T. Grover is supported by a Moore foundation fellowship under the EPiQS initiative.

\appendix

\section{Details of calculations for $\langle \rho_A \rangle$}

\subsection{Entanglement spectrum of $\langle \rho_A \rangle$} \label{sec:rhoAspect}

The reduced density matrix for the wavefunction $|\psi \rangle = \sum_C |\psi(C)| \,|C\rangle$, where $\psi(C)$ are chosen from a random ensemble, is given by

\bea
\rho_A(C_A,C'_A) & = & \frac{\sum_{C_{\overline{A}}} |\psi(C_A,C_{\overline{A}})| |\psi(C'_A, C_{\overline{A}})|} {\sum_{CA,C_{\overline{A}}} |\psi(C_A,C_{\overline{A}})|^2 } \nonumber \\
& = & \delta_{C_A,C'_A}\, \frac{\sum_{C_{\overline{A}}} |\psi(C_A,C_{\overline{A}})|^2}{\sum_{C_A,C_{\overline{A}}} |\psi(C_A,C_{\overline{A}})|^2 } +  \nonumber \\ 
& & (1- \delta_{C_A,C'_A}) \, \frac{\sum_{C_{\overline{A}}} |\psi(C_A,C_{\overline{A}})| |\psi(C'_A,C_{\overline{A}})|}{\sum_{C_A,C_{\overline{A}}} |\psi(C_A,C_{\overline{A}})|^2 } \nonumber
\eea

One may now perform an average over the random ensemble to obtain $\langle \rho_A \rangle$:

\be
\langle \rho_A\rangle = \delta_{C_A,C'_A}\frac{1}{|\mathcal{H}_A|} +  g \frac{ (1- \delta_{C_A,C'_A})}{|\mathcal{H}_A|} 
\ee
where $g = \frac{\langle |\psi(C)|\rangle^2}{\langle |\psi(C)|^2 \rangle}$ and $|\mathcal{H}_A|$ denotes the size of the Hilbert space in subregion A. The simple structure of $\langle \rho_A\rangle$ readily allows one to diagonalize it: there is a single eigenvector with eigenvalue $\lambda = g + \frac{1-g}{\mathcal{H}_A}$, and $|\mathcal{H}_A|-1$ degenerate eigenvectors with eigenvalue $\frac{1-g}{\mathcal{H}_A}$. This leads to the result for the Renyi entropies in Eqn.\ref{eq:eeavgrho0}.

The huge gap in the entanglement spectrum between the single lowest lying eigenvalue and the rest of the states leads to a finite temperature phase transition for the entanglement Hamiltonian $H_\rho = -\ln(\langle \rho_A\rangle)$ at unit temperature, as reflected in the qualitative difference between the scaling of the Renyi entropies $S_n(\langle \rho_A \rangle)$ depending on whether $n \leq 1$ (volume law), or $ n > 1$ (constant law).

\subsection{Calculation of parameter $g$ for random wavefunction ensembles} \label{sec:g}

As is evident from the discussion above, the entanglement entropies $S_n(\langle \rho_A \rangle)$ for a particular choice of ensemble depend crucially on the parameter $g$. Let us consider the two cases discussed in the main text separately:

(a) Random Positive Ensemble (RPE):

In this case $\psi(C)$ is distributed randomly and uniformly on $S^N$ where $N = 2 |\mathcal{H}_A| |\mathcal{H}_{\overline{A}}|-1$, or $|\mathcal{H}_A| |\mathcal{H}_{\overline{A}}|-1$ depending on whether the wavefunction is complex or real, where the latter case might be relevant to time-reversal invariant systems (for example).

The parameter $g$ is given by,

\be
g = \frac{\langle |\vec{\psi}| \rangle^2}{\langle |\vec{\psi}|^2 \rangle} , \label{eq:g}
\ee
where $\vec{\psi}$ is an $N$-component vector and the average is taken over a uniform distribution. We employ the following polar coordinates for our calculation:

\bea 
\psi_1 & = & \cos(\phi_1) \nonumber \\
\psi_2 & = & \sin(\phi_1) \cos(\phi_2) \nonumber \\
\psi_3 & = & \sin(\phi_1) \sin(\phi_2) \cos(\phi_3) \nonumber \\
&...& \nonumber \\
\psi_N & = & \sin(\phi_1) \sin(\phi_2)\,\, ... \,\, \sin(\phi_{N-2}) \sin(\phi_{N-1}) ,\nonumber \\
\eea
where the angles $\phi_1$ to $\phi_{N-2}$ lie between $0$ and $\pi$, while $\phi_{N-1}$ lie between $0$ and $2\pi$. To calculate the expression in \ref{eq:g}, it suffices to restrict $\phi_1$ to $\phi_{N-2}$ to the interval $\left[ 0, \pi/2 \right]$ and $\phi_{N-1}$ to $[0,\pi)$ and replace $|\psi| \rightarrow \psi$ since all the coordinates are positive within this restricted domain.

The denominator in Eqn.\ref{eq:g} is calculable trivially: $\langle |\vec{\psi}|^2 \rangle = \frac{1}{N}$ since $\sum_{i=1}^{N} |\psi_i|^2 = 1$. On the other hand, $\langle |\psi| \rangle$ is given by:

\bea
& & \langle |\psi| \rangle  \nonumber \\
& & = \frac{\int d\phi_1 \,\cos(\phi_1) \sin^{N-2}(\phi_1)}{\int d\phi_1\,\sin^{N-2}(\phi_1)} \nonumber \\
& & = \begin{cases}
\frac{1}{N-1}\frac{2}{\pi}\frac{(N-2)!!}{(N-3)!!}  \,\, \text{if  } N \,\textrm{is even},\\
\\
\frac{1}{N-1}\frac{(N-2)!!}{(N-3)!!}  \,\, \text{if  } N \,\textrm{is odd}.\\ \label{eq:avgmodpsi1}
\end{cases}
\eea
When $N \gg 1$, one may approximate the factorials in Eqn.\ref{eq:avgmodpsi1} using Sterling's formula $n! \approx \sqrt{2\pi n}\left(\frac{n}{e}\right)^n$, and one finds,

\be 
\langle |\psi| \rangle \approx \sqrt{\frac{2}{\pi N}} ,  \label{eq:avgmodpsi2}
\ee
 irrespective of whether $N$ is even or odd (as one might expect). Combining Eqn.\ref{eq:avgmodpsi2} with $\langle |\vec{\psi}|^2 \rangle = \frac{1}{N}$, one finds that $g = 2/\pi$ which leads to Eqn.\ref{eq:eeavgrho1} in the main text.

(b)  {\textit{$\psi(C)= det(M)$ where $M$ is  a matrix with random $\pm 1$ entries:}}

Recall that to obtain Renyi entropies, one requires the ratio $\frac{\langle |\det(M)|\rangle^2}{\langle |\det(M)|^2\rangle}$. Clearly, the denominator $\langle |\det(M)|^2 \rangle = L!$ where $L$ is the size of the matrix $L$ (=the total number of particles). One might have naively guessed that the numerator $= \langle |\det(M)|\rangle^2$ scales in the same fashion with $L$. However, this turns out to be incorrect. The problem of the expectation value of the modulus of a determinant was studied recently in the mathematics literature by Nguyen and Vu in Ref.\cite{nguyen2014}. They found that $\ln |\det(M)|$ is normal distributed with mean $\ln(\sqrt{(L-1)!})$, and variance $\sqrt{\frac{\ln(L)}{2}}$. From this, one finds that $\langle |\det(M)|\rangle^2 \sim (L-1)! \sqrt{L}$, and therefore, the Renyi entropies $S_n(\langle \rho_A \rangle) \sim \ln(L)$ for $n >1$, while they continue to follow a volume law for $n<1$ and for the von Neumann entropy (Eqn.\ref{eq:eeavgrho2}). As discussed in the main text, this result yields the correct scaling of $\langle S_n\rangle$ only when $\ell_A/L$ is a non-zero constant as $L \rightarrow \infty$, so that there remains only one length scale in the problem.

\section{Calculation of $ S_n(\langle \tr\,\, \rho_A^n \rangle ) $}   \label{sec:trrhoAn}

By definition, $\langle \tr \rho^n_A \rangle $ for the wavefunction $\big|\,|\psi| \,\big\rangle$ in Eq.\ref{eq:rpe} is given by,

\bea 
\langle \tr \rho^n_A \rangle & = & \sum_{\{A_i,\overline{A}_i\}} \langle |\psi(C_{A_1}\,C_{\overline{A}_1})| \,|\psi(C_{A_2}\,C_{\overline{A}_1})|\,  |\psi(C_{A_2}\,C_{\overline{A}_2})| \,\nonumber \\
& & ...|\psi(C_{A_n}\,C_{\overline{A}_n})|\, |\psi(C_{A_1}\,C_{\overline{A}_n})| \rangle . \label{eq:trrhon}
\eea
A bit of thought will convince the reader that the leading contribution to the above expression (in the limit where the size of the Hilbert space is taken to infinity), comes from the terms where all $A_i$'s and $\overline{A}_i$'s are {\it distinct}. Interestingly, such a contribution does not exist for the average over a random real or complex wavefunction \cite{lubkin1978}. This is because, if the wavefunction was allowed to take both positive and negative values, and if all  $A_i$,$\overline{A}_i$ are distinct, the contributions cancel out pairwise. This crucial difference  leads to a qualitatively different behavior of entanglement entropy $ S_n(\langle \tr\,\, \rho_A^n \rangle ) $ for a positive wavefunction.

The above average, to the leading order, is

\bea 
\langle \tr \,\,\rho^n_A \rangle & \approx & \prod_{i=0}^{n-1} (|\mathcal{H}_A|-i)(|\mathcal{H}_{\overline{A}}|-i) \times \langle \prod_{i=1}^{2n} x_i \rangle,
\eea
  
 \noindent where $x_i$ are the first $2n$ Cartesian coordinates of the Euclidean embedding of   the unit sphere $S^{|\mathcal{H}_A||\mathcal{H}_{\overline{A}}|-1}$, and we have assumed that the wavefunction $\psi$ is real. The combinatorial prefactor multiplying   $ \langle \prod_{i=1}^{2n} x_i \rangle$ can be obtained by imposing the constraint on the expression in Eqn.\ref{eq:trrhon} that all the configurations are distinct. $ \langle \prod_{i=1}^{2n} x_i \rangle$ can be calculated conveniently via spherical polar coordinates. One finds:
 
 \be 
  \langle \prod_{i=1}^{2n} x_i \rangle =  \left( \frac{2}{\pi}\right)^n \left[ \prod_{i=1}^{n} \left( |\mathcal{H}_A| |\mathcal{H}_{\overline{A}}| + 2n - 2i\right) \right]^{-1}  .
 \ee
 
Putting everything together, and taking the limit $|\mathcal{H}_A|, |\mathcal{H}_{\overline{A}}| \gg n$, one finds that at the leading order,
 
 \be 
 \langle \tr\,\, \rho^n_A \rangle \approx \left(\frac{2}{\pi} \right)^n  .
 \ee
 
Therefore,  $ S_n(\langle \tr\,\, \rho_A^n \rangle ) $ for $n$ integer ($n>1$) is given by,
 
 \be 
 S_n(\langle \tr\,\, \rho_A^n \rangle ) = \frac{n}{n-1} \ln\left(\frac{\pi}{2}\right) ,\label{eq:Savgtrrhon}
 \ee 
which precisely matches the results for  $S_n(\langle  \rho_A \rangle )$ and  $\langle S_n(\rho_A)\rangle$ discussed in the main text. Even though we only discussed the case where $\psi$ is real, the above calculation trivially generalizes to the case when $\psi$ is complex, and the answer (Eq.\ref{eq:Savgtrrhon}) remains unchanged.


\begin{thebibliography}{x}
\bibitem{footnote:sign} The distinction between real wavefunctions (i.e. wavefunctions which take both positive and negative values, but are not complex) and complex wavefunctions turns out to be irrelevant for our discussion, both of whom thus possessing a non-trivial sign structure in our nomenclature.
\end{thebibliography}
\end{document}